\documentclass[useAMS,usenatbib]{mn2e}
\usepackage{graphicx}
\usepackage{times}
\usepackage{amssymb}
\usepackage{amsmath}
\begin{document}

\title[4\% Distance Measure at $z=0.15$] 
{The Clustering of the SDSS DR7 Main Galaxy Sample I: A 4 per cent Distance Measure at $z=0.15$}

\author[A. J. Ross et al.]{\parbox{\textwidth}{
Ashley J. Ross\thanks{Email: Ashley.Ross@port.ac.uk, ross.1333@osu.edu}$^{1,2}$, 
Lado Samushia$^{1,3,4}$,
Cullan Howlett$^{1}$,
Will J. Percival$^{1}$,
Angela Burden$^{1}$,
Marc Manera$^{5,1}$
}
  \vspace*{4pt} \\ 
$^{1}$Institute of Cosmology \& Gravitation, Dennis Sciama Building, University of Portsmouth, Portsmouth, PO1 3FX, UK\\
$^{2}$Center for Cosmology and AstroParticle Physics, The Ohio State University, Columbus, OH 43210, USA\\
$^{3}$Department of Physics, Kansas State University, 116, Cardwell Hall, Manhattan, KS, 66506, USA\\
$^{4}$National Abastumani Astrophysical Observatory, Ilia State University, 2A Kazbegi Ave., GE-1060 Tbilisi, Georgia\\
$^{5}$University College London, Gower Street, London WC1E 6BT, UK\\
}
\date{Accepted by MNRAS}

\pagerange{\pageref{firstpage}--\pageref{lastpage}} \pubyear{2014}
\maketitle
\label{firstpage}

\begin{abstract}
We create a sample of spectroscopically identified galaxies with $z < 0.2$ from the Sloan Digital Sky Survey (SDSS) Data Release 7, covering 6813 deg$^2$. Galaxies are chosen to sample the highest mass haloes, with an effective bias of 1.5, allowing us to construct 1000 mock galaxy catalogs (described in Paper II), which we use to estimate statistical errors and test our methods. We use an estimate of the gravitational potential to ``reconstruct'' the linear density fluctuations, enhancing the Baryon Acoustic Oscillation (BAO) signal in the measured correlation function and power spectrum. Fitting to these measurements, we determine $D_{V}(z_{\rm eff}=0.15) =  (664\pm25)(r_d/r_{d,{\rm fid}})$ Mpc; this is a better than 4 per cent distance measurement. This ``fills the gap'' in BAO distance ladder between previously measured local and higher redshift measurements, and affords significant improvement in constraining the properties of dark energy. Combining our measurement with other BAO measurements from BOSS and 6dFGS galaxy samples provides a 15 per cent improvement in the determination of the equation of state of dark energy and the value of the Hubble parameter at $z=0$ ($H_0$). Our measurement is fully consistent with the Planck results and the $\Lambda$CDM concordance cosmology, but increases the tension between Planck$+$BAO $H_0$ determinations and direct $H_0$ measurements.
\end{abstract}


\section{Introduction}
\label{sec:intro}
Robust measurements of the cosmological expansion rate are required to understand its observed acceleration at low redshift (see \citealt{riess98, Perl99} for early detections and \citealt{Weinberg12} for review of observational probes). If the acceleration is driven by an unknown energy-density component, then these measurements will constrain its equation-of-state. If the effect is instead a manifestation of large-scale gravity that is different from General Relativity, then we can test for this by combining these measurements with observations of cosmological structure growth. Galaxy surveys can provide measurements of both through Baryon Acoustic Oscillations (BAO), which act as a fixed ruler allowing the expansion rate to be determined, and Redshift-Space Distortions (RSD), which are imprints of the velocity field that allow the rate of structure growth to be determined. Ideally we need accurate measurements at a series of redshift bins, building a full understanding of cosmological acceleration.

The Sloan Digital Sky Survey (SDSS; \citealt{York00}) data release seven (DR7; \citealt{DR7}) included a flux-limited, low-redshift sample of galaxies with measured redshifts, known as the `main galaxy sample' (MGS). Baryon Acoustic Oscillations have previously been measured from these data, when included together with higher redshift samples (e.g. \citealt{Per10}). However, these previous analyses did not utilise recent analysis developments that use phase information to ``reconstruct'' the linear fluctuations on scales important for BAO measurements. Such techniques can provide a $\sim$50\% improvement on the measured BAO scale \citep{Eis07rec,Pad09,Burden} and the improvement is expected to be larger at lower redshifts. In this paper, we will analyse these data using an up-to-date pipeline similar to that used by \cite{alphdr11} for the recent SDSS-III \citep{Eis11} Baryon Oscillation Spectroscopic Survey (BOSS; \citealt{Daw12}) and including the reconstruction technique, which has proven to improve the precision of BAO measurements in multiple sets of galaxy redshift survey data \citep{Xu12,Ross14,alphdr11,Toj14,Kazin14}.

By performing this analysis we hope to complete the set of accurate BAO-distance measurements that can be made from current data using the reconstruction technique. The SDSS data covers a larger area than the 2-degree Field Galaxy Redshift Survey (2dFGRS; \citealt{colless03}), and will thus provide more accurate measurements at similar redshifts \citep{per01,cole05}. This analysis builds on a rich recent history of BAO measurements \citep{Per10,Blake11,Pad12,Kazin14}, and complements measurements at lower redshift made by \cite{Beutler11} using data from the 6-degree Field Galaxy Redshift Survey (6dFGS; \citealt{Jones09}), and at higher redshift by \cite{Toj14} and \cite{alphdr11} using BOSS data.

BOSS measures redshifts for galaxy samples selected using two different algorithms, a redshift $z\sim0.57$ sample known as CMASS (which was selected to a approximately constant stellar mass threshold) and a sample at $z\sim0.32$ known as LOWZ. While these measurements are more precise than the ones obtainable from MGS data, the MGS samples an independent cosmic volume and a lower redshift range, which is important when studying properties of dark energy since its effects are more pronounced at lower redshifts.

We apply colour, magnitude, and redshift cuts to the SDSS DR7 MGS data to produce a cosmic-variance limited spectroscopic sample with $z<0.2$, where the galaxies inhabit high-mass dark matter halos. This ensures that we can easily simulate the catalogue with mock galaxy catalogs to test our methods and provide covariance matrices. We have produced $1000$ accurate mock galaxy catalogs, based on fast numerical N-body simulations \citep{Howlett14a}. A Halo-Occupation Distribution (HOD) model for the galaxy distribution was applied to the results from fast numerical N-body simulations as described in a companion paper \citep{Howlett14b}, which also presents RSD measurements made with this sample. 

In this paper, we present the sample (Section~\ref{sec:data}), BAO analysis (Section~\ref{sec:analysis}), tests on the mocks (Section~\ref{sec:mocks}), results (Section~\ref{sec:results}), and cosmological interpretation (Section~\ref{sec:cosmology}). Throughout, we assume a fiducial cosmology given by $\Omega_m=0.31$, $\Omega_b=0.048$, $h=0.67$, $\sigma_8=0.83$, $n_s=0.96$, and $\Omega_{\nu}=0$ which matches that used to create the mock catalogues, and is based on the best-fit $\Lambda$CDM model of \cite{alphdr11}, and is consistent with Planck satellite \citep{PlanckOver} cosmic microwave background (CMB) data \citep{PlanckCos}.

\section{Data}  \label{sec:data}

The SDSS DR7 contains the completed data set of SDSS-I and SDSS-II. These surveys obtained wide-field CCD photometry (\citealt{C,Gunn06}) in five passbands ($u,g,r,i,z$; \citealt{F}), internally calibrated using the `uber-calibration' process described in \cite{Pad08}, amassing a total footprint of 11,663 deg$^2$. From this imaging data, galaxies within a footprint of 9380 deg$^2$ \citep{DR7} were selected for spectroscopic follow-up as part of the main galaxy sample (MGS; \citealt{Strauss02}), which, to good approximation, consists of all galaxies with $r_{\rm pet} < 17.77$, where $r_{\rm pet}$ is the extinction-corrected $r$-band Petrosian magnitude. In this analysis we do not consider the Luminous Red Galaxy extension of this program to higher redshift \citep{Eis01}.

We obtain the SDSS DR7 MGS data from the value-added galaxy catalogs hosted by NYU\footnote{http://sdss.physics.nyu.edu/vagc/lss.html} (NYU-VAGC). These catalogs were created following the methods described in \cite{Blanton05}. They include $K$-corrected absolute magnitudes, determined using the methods of \cite{Blanton03}, and detailed information on the mask. We select our galaxy sample from the NYU-VAGC `safe0' catalog. This sample uses galaxies with $14.5 < r_{\rm pet} < 17.6$. The $r_{\rm pet} > 14.5$ limit ensures that only galaxies with reliable SDSS photometry are used and the $r_{\rm pet} < 17.6$ allows a homogeneous selection over the full footprint of 7356 deg$^2$ \citep{Blanton05}. Galaxies that did not obtain a redshift due to fibre collisions are given the redshift of their nearest neighbour.

\begin{figure}
\includegraphics[width=84mm]{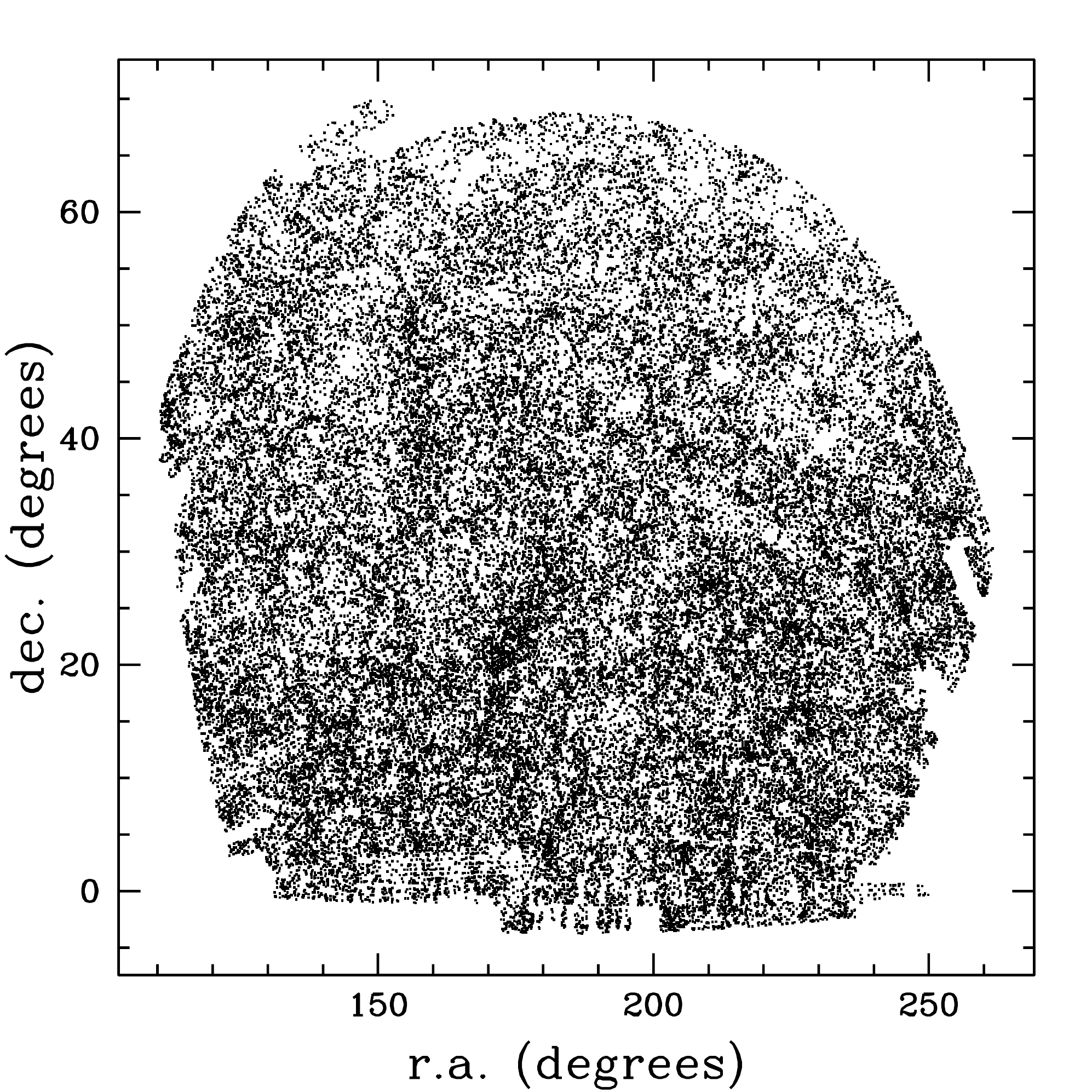}
  \caption{The right ascension and declination positions (J2000) of the 63,163 SDSS DR7 main galaxy survey galaxies we include in our sample. Their footprint occupies 6813 deg$^2$.}
  \label{fig:foot}
\end{figure}

Galaxies in the NYU VAGC safe0 catalog occupy a total footprint of 7356 deg$^2$. We use data only from the contiguous area in the North Galactic cap and only occupying areas where the completeness, determined ignoring galaxies not observed due to fiber collisions, is greater than 0.9. These cuts reduce the footprint to 6813 deg$^2$. The angular positions of the galaxy sample we use are plotted in Fig.~\ref{fig:foot}. In order to obtain angular positions for random catalogs, we use the window given by the NYU-VAGC and the {\sc Mangle} software \citep{mangle}, down-sampling based on the completeness in each region (as provided in the window).

We make further cuts on the NYU VAGC safe0 sample based on colour, magnitude, and redshift to produce our catalog, which we refer to as MGS from here on. These cuts balance the following motivations: 
\begin{enumerate}
  \item Create a sample that is at a lower redshift than is probed by BOSS. We therefore use only galaxies with $z < 0.2$
  \item Reliably simulate the clustering of the galaxies in our sample. In order to do so, we require a reasonably constant galaxy density as a function of redshift $n(z)$ and that galaxies occupy dark matter halos with masses $M_{\rm halo} > 10^{12}\,M_\odot$, which is the minimum halo mass that our simulations can reliably achieve.
  \item Minimize the fractional uncertainty expected for measurements of $P(k)$, balanced against the above two concerns. This is achieved by maximizing the galaxy density.
\end{enumerate}
Balancing these motivations, we define our sample to have $0.07 < z < 0.2$, $M_r < -21.2$ and $g-r > 0.8$, where $M_r$ is the $r$-band absolute magnitude provided by the NYU-VAGC. The resulting sample contains 63,163 galaxies. The luminosity and colour cuts make the sample more homogenous as a function of redshift and increase the clustering amplitude of the sample. The increase in clustering amplitude implies an increase in the mass of the typical halo hosting one of our sample galaxies. The results presented in \cite{Zeh11} (e.g., their figure 10), suggest a negligible fraction of galaxies matching our selection occupy halos with $M < 10^{12}\,M_\odot$ and therefore imply we will be able to reliably simulate our sample. The $z > 0.07$ limit is applied as, due to the $r_{\rm pet} > 14.5$ cut, the number density of our sample drops sharply for $z< 0.07$.

\begin{figure}
\includegraphics[width=84mm]{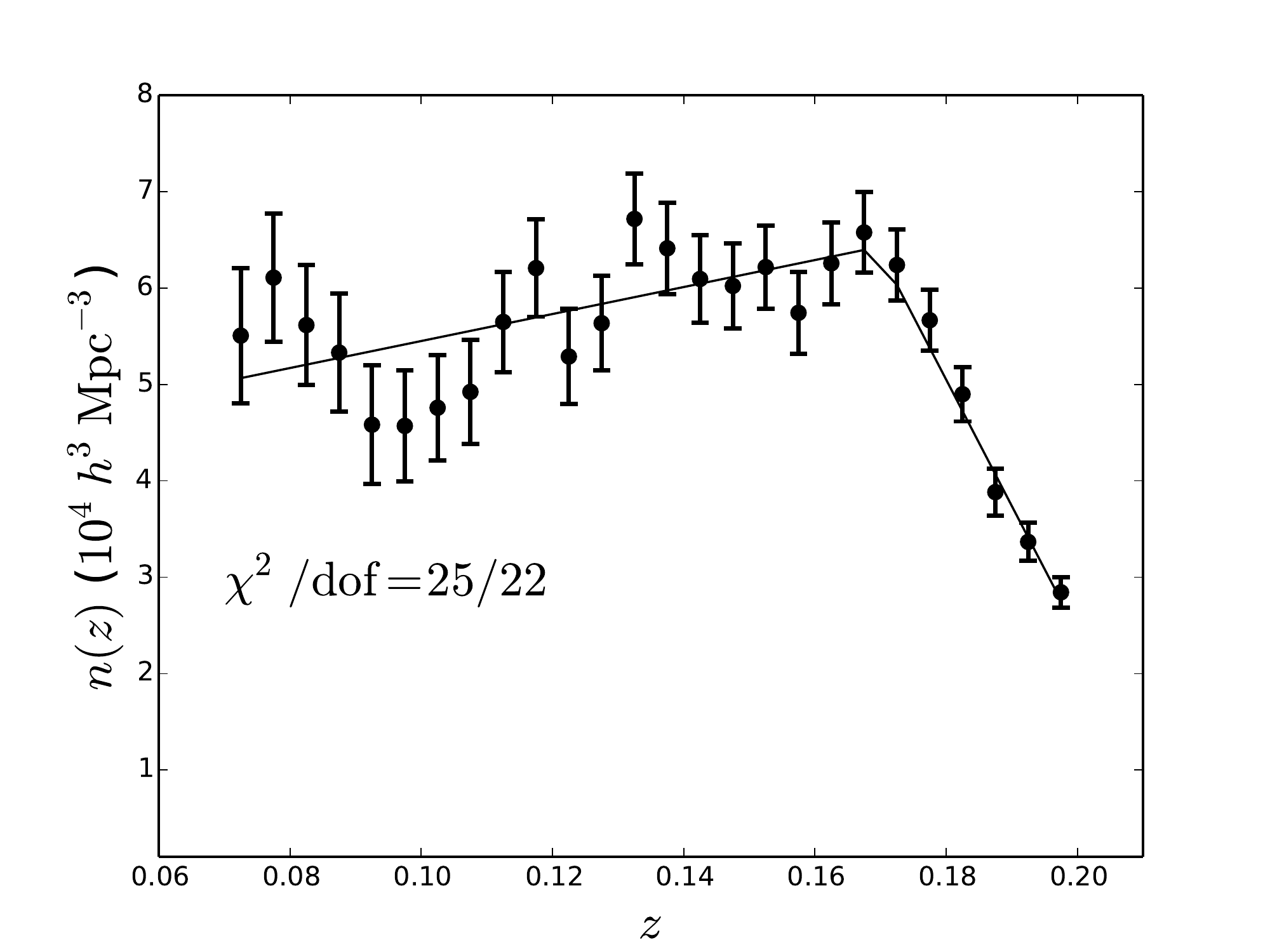}
  \caption{The number density of the sample we use plotted as function of redshift. The error-bars represent the standard deviation of $n(z)$ for the mock realizations. The curve is the best-fit model assuming two linear relationships with a transition redshift, which is a good fit to the data.}
  \label{fig:nz}
\end{figure}
 
The $n(z)$ of our sample is displayed in Fig.~\ref{fig:nz}. We fit this $n(z)$ to a model with two linear relationships and a transition redshift, given by
\begin{eqnarray}
& n(z) = 0.0014z+0.00041; z < 0.17 \label{eq:nz}\\ 
& n(z) = 0.00286-0.0131z; z\geq 0.17,\label{eq:nz2}
\end{eqnarray}
which provides a good representation of the data, as the $\chi^2$ is 25 for 22 degrees of freedom (26 $n(z)$ bins and four independent model parameters). This function is used to create the mock catalogs (see Section~\ref{sec:mocks}), when assigning redshift-dependent $w_{\rm FKP}$ weights for the clustering measurements (see Eq.~\ref{eq:wfkp}), and when assigning redshifts to the random catalog we use to measure the clustering of mock samples. We do not use the analytic fit to assign redshifts to the angular positions in the random catalog we use to measure the clustering of our data sample; instead we randomly select redshifts from the galaxy catalog, thus allowing for any further observation-dependent fluctuations. This procedure was shown to impart negligible bias on BOSS clustering measurements in \cite{DR9sys}, and in Paper II \citep{Howlett14b} it is shown that it makes a negligible difference for our MGS sample. 

Our galaxy sample is approximately volume limited to $z < 0.17$ (due to the $M_r < -21.2$ restriction), above which the number density drops due to the $r_{\rm pet} < 17.6$ magnitude limit. Speculatively, the slight increase in number density with redshift may be due to evolution in the stellar populations of passive galaxies; as the stellar populations of the galaxies age, many will dim and may drop out of our sample. We have not attempted to correct for such effects. The $z > 0.17$ data is important, as it represents a larger fraction of the volume of our sample. The $n(z)$ is large enough that the sample is cosmic-variance limited ($n(z)P(k) > 1$) over the entire redshift range for $k < 0.26h$Mpc$^{-1}$. Further, the $n(z)$ is constant to within a factor of two, making it more constant than the BOSS `CMASS' sample that has been modelled as having a single HOD at its effective redshift in cosmological analyses (e.g., \citealt{alph,alphdr11}). The effective redshift of our sample is $z_{\rm eff} = 0.15$, calculated using $z_{\rm eff} = \frac{\int dV_{\rm eff} z }{\int dV_{\rm eff}}$, where $dV_{\rm eff}$ is \citep{Tegmark98} $dV_{\rm eff} = dV\left[n(z)P_{\rm FKP}w_{\rm FKP}\right]^2$, with $P_{\rm FKP}$ and $w_{\rm FKP}$ defined in the following section.

\begin{figure*}
\centering
\begin{minipage}{5in}
\includegraphics[width=5in]{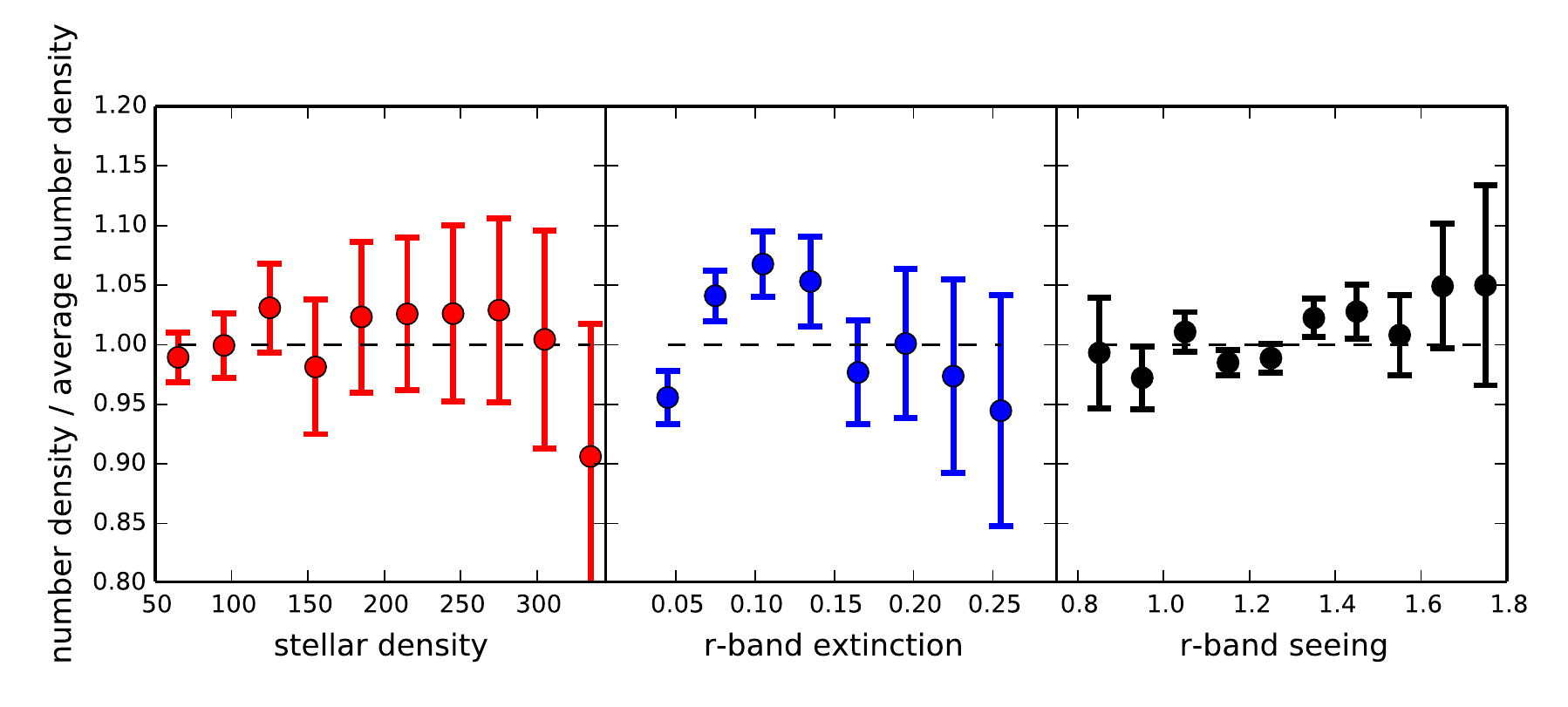}
  \caption{The number density of our galaxy sample as a function of stellar density, Galactic extinction, and seeing. The error-bars denote the standard deviation found in the mock galaxy samples and are only due to stochastic variations in the galaxy field. The variations in number density are consistent with the expected level of fluctuation.}
  \label{fig:nsys}
 \end{minipage} 
\end{figure*}

Because of the depth of the target sample from which galaxies were selected for spectroscopic follow-up in the BOSS CMASS galaxy sample, some angular fluctuations in the catalogue were imparted by changes in observational parameters. This  non-cosmological clustering signal was corrected using ``systematic weights'' \citep{DR9sys}. Such a correction was not required for the brighter BOSS LOWZ sample \citep{Toj14}. Fig~\ref{fig:nsys} displays the number density of our MGS data against stellar density, extinction and seeing. No significant systematic relationships are found. The variations with seeing and stellar are consistent with the expected level of variation. For extinction, the variation is somewhat larger than expected (the $\chi^2$/dof = 16.4 for eight measurement bins when the data is compared to a model with no variation), but the data exhibits no clear trend. That we find no evidence of observational systematics is not a surprise, as the sample we use is approximately two magnitudes brighter than the LOWZ sample, and we therefore do not expect systematic weights to be necessary.

\section{Analysis}  \label{sec:analysis}

\subsection{Calculating Clustering Statistics}

We calculate the correlation function as a function of the redshift-space separation $s$ and the cosine of the angle to the line-of-sight, $\mu$,
 using the standard \cite{LS} method 
\begin{equation}
\xi(s,\mu) =\frac{DD(s,\mu)-2DR(s,\mu)+RR(s,\mu)}{RR(s,\mu)}, 
\label{eq:xicalc}
\end{equation}
where $D$ represents the galaxy sample and $R$ represents the uniform random sample that simulates the selection function of the galaxies. $DD(s,\mu)$ thus represent the number of pairs of galaxies with separation $s$ and orientation $\mu$. We use at least fifty times the number of galaxies in our random samples, both for the mock realisations and the DR7 data. 

We calculate $\xi(s,|\mu|)$ in evenly-spaced bins\footnote{The pair-counts are tabulated using a bin width of 1 $h^{-1}$Mpc and then summed into 8 $h^{-1}$Mpc bins, allowing different choices for bin centres .} of width 8 $h^{-1}$Mpc in $s$ and 0.01 in $|\mu|$. We then determine the first two even moments of the redshift-space correlation function via 
\begin{equation}
\frac{2\xi_{\ell}(s)}{2\ell+1} = \sum^{100}_{i=1} 0.01\xi(s,\mu_i)L_{\ell}(\mu_i), 
\label{eq:xiell}
\end{equation}
where $\mu_i = 0.01i-0.005$ and $L_\ell$ is a Legendre polynomial of order $\ell$. We will only use $\xi_0$ in this paper (an analysis including aniostropic information is presented in \citealt{Howlett14b}). Defining $\xi_0$ as we do ensures equal weighting in $\mu$, which is not necessarily the case when the calculations are not split into $\mu$ bins. Defining $\xi_0$ in this fashion forces the measurement to be closer to a spherical average and thus ensures BAO distance measurements can be expressed as $D_{V}$, defined in Eq. \ref{eq:Dv}.

We weight both galaxies and randoms based on the number density as a function of redshift \citep{FKP}, via
\begin{equation}
w_{\rm FKP}(z) = \frac{1}{1+P_{\rm FKP}n(z)},
\label{eq:wfkp}
\end{equation}
where we set $P_{\rm FKP} = 16000 h^{-3}$Mpc$^{3}$, which is close the the measured amplitude at $k=0.1 h$Mpc$^{-1}$. 

We measure the spherically-averaged power spectrum, $P(k)$, using the standard Fourier technique of \cite{FKP}, as described in \cite{ReidDR7} and \cite{alph}. We calculate the spherically-averaged power in $k$ bands of width $\Delta k = 0.008h$Mpc$^{-1}$ using a 1024$^3$ grid. The weights are taken into account by using the sum of $w_{\rm FKP}$ over the galaxies/randoms at each gridpoint.

 \subsection{Reconstruction}

Reconstruction of galaxy clustering data (\citealt{Eis07rec}) has now been shown to improve measurements of the BAO scale in multiple galaxy samples, including SDSS-II LRGs at $z=0.35$ (\citealt{Pad12} and \citealt{Xu13}), the SDSS-III BOSS LOWZ and CMASS samples (\citealt{alphdr11} and \citealt{Toj14}), red and blue galaxies in the CMASS sample \citep{Ross14} and emission line galaxies from the WiggleZ survey \citep{Kazin14}. Reconstruction uses the galaxy map to construct a displacement field that is used to redistribute the galaxies into a spatial configuration that more closely reproduces their positions had they only undergone linear growth and removes the effect of redshift space distortions. The process thereby (typically) sharpens the BAO feature in clustering measurements and removes non-linear shifts in its peak position, thus allowing significantly more precise and accurate BAO measurements.

The algorithm used in this paper is similar to the prescription of \cite{Eis07rec} and \cite{Pad12} and is the same as applied in \cite{Toj14}.
The Lagrangian displacement field $\bold{\Psi}$ is calculated to first order using the Zel'dovich approximation applied to the smoothed galaxy overdensity field. The displacement field is corrected for redshift effects in the measured overdensity. Our implementation deviates from \cite{Pad12} slightly in that we solve for the redshift-space-corrected displacement field in Fourier space rather than using the finite difference method in configuration space, although we find both methods in very good agreement, as shown in \cite{Burden}. Further details can be found in \cite{Pad12} and \cite{Burden}. We use a bias value of $b=1.5$, a linear growth rate of $f=0.6413$ and a smoothing scale of 15$h^{-1}$ Mpc. The bias value is close to the best-fit bias found by \cite{Howlett14b}. Previous studies, e.g., \cite{Pad12, alph,alphdr11}, have found results to be insensitive to the choice of bias used in the reconstruction implementation.

We present clustering measurements and covariances using both original and reconstructed catalogues. Following \cite{Pad12}, the reconstructed $\xi$ is calculated as
\begin{equation}
\xi(s,\mu) =\frac{DD(s,\mu)-2DS(s,\mu)+SS(s,\mu)}{RR(s,\mu)}, 
\label{eq:xicalcrec}
\end{equation}
where $S$ represents the shifted random field. The $P(k)$ is calculated using the shifted random field, but models are compared to the measurement using the window function determined from the original random field.

\subsection{Measuring BAO Positions}

The methodology we use to measure isotropic BAO positions is adapted from and nearly identical to \cite{alphdr11} and \cite{Toj14}. We repeat the basic details here and identify key differences but refer to \cite{alphdr11} for more detailed descriptions.

In order to measure BAO positions, we extract a dilation factor $\alpha$ by comparing our data to a template that includes the BAO and a smooth curve with considerable freedom in its shape that we marginalise over. For fits to the monopole, assuming pairs at all angles to the line-of-sight contribute equally, measurements of $\alpha$ can be related to physical distances via
\begin{equation}
\alpha = \frac{D_V(z)r_d^{\rm fid}}{D^{\rm fid}_V(z)r_d}
\label{eq:alpha}
\end{equation}
where
\begin{equation}
D_V(z) \equiv \left[cz(1+z)^2D^2_A(z)H^{-1}(z)\right]^{1/3},
\label{eq:Dv}
\end{equation}
$r_d$ is the sound horizon at the baryon drag epoch, which can be accurately calculated for a given cosmology using, e.g., the software package {\sc CAMB} \citep{Lewis00}, $D_A(z)$ is the angular diameter distance, and $H(z)$ is the Hubble parameter. For our fiducial cosmology, $D^{\rm fid}_V(z) = 638.95$ Mpc and $r_d^{\rm fid} = 148.69$ Mpc.

We fit the measured, spherically averaged, correlation function and
power spectrum separately and then combine results using the mocks to
quantify the correlation coefficient between measurements. Our fits rescale
a model of the damped BAO in order to fit the data. Broad-band effects, caused by, e.g.,
errors made in our assumption of the model cosmology, scale-dependent bias, and redshift-space distortions, are
marginalised over using polynomial terms. 

For both $P(k)$ and $\xi(s)$, we use the linear theory $P(k)$ produced by {\sc CAMB} and split it into a smooth ``De-Wiggled'' template
$P^{\rm sm, lin}$ and a BAO template ${\rm O}^{\rm lin}$,
following \citet{EisSeoWhi07} and using the fitting formulae of \cite{EH98}. The damped BAO feature is then given by
\begin{equation}
{\rm O^{damp}}(k) = 1+({\rm O^{lin}}(k)-1)e^{-\frac{1}{2}k^2\Sigma^2_{nl}}
\end{equation} 

For the $P(k)$ fits, the damping is treated as a free parameter, with a Gaussian prior of width $\pm 2h^{-1}$Mpc centred at the best-fit value recovered from the mocks. Pre-reconstruction this is $\Sigma_{nl} = 9.0 h^{-1}$Mpc and post-reconstruction it is $\Sigma_{nl} = 5.3 h^{-1}$Mpc. The full model fitted for $P(k)$ is
\begin{equation}  \label{eq:mod_pk}
  P^{\rm fit}(k)=P^{\rm sm}(k){\rm O^{damp}}(k/\alpha), 
\end{equation}
where 
\begin{equation} \label{eq:pksm}   
    P^{\rm sm}(k)= B_p^2P(k)^{\rm sm,lin}+A_1k+A_2+\frac{A_3}{k}+\frac{A_4}{k^2}+\frac{A_5}{k^3},
\end{equation} 
These are therefore six ``nuisance'' parameters: a multiplicative
constant for an unknown large-scale bias $B_p$, and five polynomial
parameters $A_1$, $A_2$, $A_3$, $A_4$, and $A_5$. 

For the correlation function, we use a
model
\begin{equation} \label{eqn:fform}
  \xi^{\rm fit}(s) = B_\xi^2\xi^{\rm lin,damp}(\alpha s)+A_\xi(s).
\end{equation}
where $\xi^{\rm lin,damp}(s)$ is the Fourier transform of
 $P^{\rm sm, lin}(k) {\rm O^{damp}}(k)$. $B_\xi$ is a multiplicative
constant allowing for an unknown large-scale bias, while the
additive polynomial is given by
\begin{equation}
  A^\xi(s) = \frac{a_1}{s^2} + \frac{a_2}{s} + a_3
\end{equation}
where $a_i$, $1<i<3$ help marginalize over the broadband signal. 

Unlike for $P(k)$,
we do not allow the damping parameter to vary when fitting $\xi(s)$ and instead fix it at the mean best-fit
value recovered from the mocks (we test this choice in Section~\ref{sec:results}); for the reconstructed $\xi(s)$ this is $\Sigma_{nl} = 4.5 h^{-1}$Mpc. It is smaller than the best-fit value found for $P(k)$ due to the fact that the BAO feature is multiplied by the polynomial term for $P(k)$, but not for $\xi(s)$. In the $\xi(s)$ fits, it is not possible to isolate the BAO feature in a manner analogous to the ${\rm O^{damp}}(k/\alpha)$ term in the $P(k)$ fits. Thus the size of the BAO relative to $A^\xi(s)$ varies, while in the $P(k)$ fits, the size of the BAO feature is always fixed
relative to $P^{\rm sm}$. Thus the amplitude of the BAO feature has more freedom in the $\xi(s)$ model, independent of the damping parameter, and this freedom has been found to be equivalent to the damping prior we place on the $P(k)$ fits in previous studies (e.g., \citealt{alphdr11}).

The scale dilation parameter, $\alpha$, measures the relative position
of the acoustic peak in the data versus the model, thereby
characterising any observed shift.  If $\alpha>1$, the acoustic peak
is shifted towards smaller scales. For fits to both the correlation
function and power spectrum, we obtain the best-fit value of $\alpha$
assuming that $\xi(s)$ and $\log P(k)$ were drawn from multi-variate
Gaussian distributions, calculating $\chi^2$ at intervals of
$\Delta\alpha=0.001$ in the range $0.8<\alpha<1.2$. 

\subsection{Covariance}

The estimated covariance
$\tilde{\sf C}$ between statistic $X$ in measurement bin $i$ and statistic $Y$ in
measurement bin $j$ is 
\begin{equation}
\tilde{\sf C}(X_i,Y_j) = \frac{1}{N_{\rm mocks}-1}\sum_{m=1}^{N_{\rm mocks}}(X_{i,m}-\bar{X}_i)(Y_{j,m}-\bar{Y}_j),
\label{eq:cov}
\end{equation}
where the index $m$ represents a different realisation of our sample, created using the methods described in Section \ref{sec:mocks}.  Error bars shown in all plots show the square-root of the diagonal elements of these covariance matrices.

To obtain an unbiased estimate of the inverse covariance matrix $\textbf{{\sf C}}^{-1}$ we rescale
the inverse of our covariance matrix by a factor that depends on the number of
mocks and measurement bins (see e.g., \citealt{Hartlap07}) 
\begin{equation} \textbf{\sf C}^{-1} =
  \frac{N_{\rm mocks}-N_{\rm bins}-2}{N_{\rm mocks}-1}~\tilde{\textbf{\sf C}}^{-1}.  \label{eq:cinv}
\end{equation}
$N_{\rm mocks}$ is 1000 in all cases, but $N_{\rm bins}$ will change depending on the
specific test we perform. We determine $\chi^2$ statistics in the standard manner,
i.e.,  
\begin{equation}
\chi^2 = (\textbf{\it X}-\textbf{\it X}_{\rm mod}) \textbf{\sf C}_{X}^{-1} (\textbf{\it X}-\textbf{\it X}_{\rm mod})^{T},
\end{equation}
where the data/model vector $\textbf{{\it X}}$ can contain any combination of clustering measurements. Likelihood distributions, ${\cal L}$, are determined by assuming ${\cal L}(\textbf{\it X}) \propto e^{-\chi^2(X)/2}$. 

\cite{Per14} derived factors that correct uncertainties determined using a covariance matrix that is constructed from mock realisations and standard deviations determined from those realisations so that they account for biases that result from the finite number of realisations employed. Defining
\begin{equation}
A = \frac{1}{(N_{\rm mocks}-N_{\rm bins}-1)(N_{\rm mocks}-N_{\rm bins}-4)},
\end{equation}
and
\begin{equation}
B = A\left(N_{\rm mocks}-N_{\rm bins}-2\right),
\end{equation}
the variance estimated from the likelihood distribution should be multiplied by
\begin{equation}
m_{\sigma} = \frac{1+B(N_{\rm bins}-N_p)}{1+2A+B(N_p+1)},
\end{equation}
and the sample variance should be multiplied by
\begin{equation}
m_{v} = m_{\sigma}\frac{N_{\rm mocks}-1}{N_{\rm mocks}-N_{\rm bins}-2}.
\end{equation}
We apply these factors, where appropriate, to all values we quote. The corrections are rather small, because we use 1000 mock realisations. This makes the corrective factors to the recovered uncertainty and standard deviations $\sqrt{m_{\sigma}} < 1.01$ and $\sqrt{m_v} < 1.03$. 

\section{Mock samples}
\label{sec:mocks}
\subsection{Creating Mock Samples}
In order to estimate the covariance matrix for our measured correlation function and power spectrum we have generated a sample of 1000 mock catalogues.  In order to produce the number of mocks required to recover accurate covariance matrices, some realism must be sacrificed. Our mocks are based on a single redshift output ($z=0.15$) and are based on dark matter simulations, from which halo catalogs are generated and populated with galaxies. We summarise the process below, full details can be found in \cite{Howlett14a,Howlett14b}.

We begin by generating 500 independent dark matter fields at a redshift of z=0.15, based on our fiducial cosmology (see Section~\ref{sec:intro}). These were created using a newly-developed code {\sc PICOLA}, a highly-developed, planar-parallelised implementation of the {\sc COLA} method of \cite{Tassev13}. The method combines second order lagrangian perturbation theory (2LPT) and a Particle-Mesh N-Body algorithm to produce simulations accurate to much smaller scales than 2LPT can reach alone, but several times faster than the Particle-Mesh algorithm to reach the same level of accuracy, and is described in \citet{Howlett14a}. A similar method has also recently been used to successfully create mock catalogues for the WiggleZ survey \citep{Kazin14}, though the code was developed independently.

Simulations were constructed in a box of side length $1280h^{-1} \text{Mpc}$, which is large enough to cover the full sky out to the maximum redshift of our sample, with $1536^{3}$ particles, giving a mass resolution of $4.98\times10^{10}\,\text{M}_{\odot}h^{-1}$. Halos were identified from the evolved dark matter field using a Friends-of-Friends algorithm \citep{Davis85} with linking length $b=0.2l$ where $l$ is the mean particle separation and a halo was defined as containing at least 10 bound particles. The halos were populated using a halo occupation distribution (HOD; \citealt{PS00}) with the same five parameter functional form as that in \cite{Zheng07} and \cite{Manera12}. We find the best-fit HOD parameters by minimizing the $\chi^2$ difference between the observed galaxy power spectrum and the power spectrum recovered when we directly apply the HOD to $10$ of our mock halo catalogues. For this process we used theoretical errors on the power spectrum derived from \cite{Tegmark98}. Further details, including the HOD parameters, can be found in Paper II. 

The survey mask is applied to the 500 independent full-sky galaxy fields to select volumes matching our sample. We are able to fit the survey in the full-sky simulation twice without overlap, using the transformation $RA \to RA+180.0$ and $DEC \to -DEC$ to go from one projection to the other. This allows us to double the number of mock catalogues at the expense of some large scale covariance between mocks generated from the same dark matter field. The nearest distance between any pair of mock galaxies drawn from different mock realisations but within the same full-sky field is 170 $h^{-1}$Mpc. This is within the range of scales we fit, but we have tested and found that the correlations between the $P(k)$ calculated for different mock realisations drawn from the same full-sky field are consistent with zero for $0.02 < k < 0.3h$Mpc$^{-1}$ (the range of scales we fit for BAO measurements) and we can therefore treat them as independent.

The final step is to subsample the number density of the mock galaxy catalogues to mimic the idealised redshift-dependent number density derived from the galaxy sample. This ensures that we capture survey effects such as the magnitude limit of our sample, which artificially changes the number density compared to that predicted by our HOD model, which is a constant $7\times10^{-4} h^3$Mpc$^{-3}$ over our redshift range. We subsample by defining the probability to keep a galaxy at a given redshift as the ratio of the average number density at that redshift, over all 1000 mocks, to the fitted number density given by Eqs. \ref{eq:nz} and \ref{eq:nz2}.

Paper II shows that the clustering of the mock samples is an excellent match to the observed clustering of our MGS sample, as the $\chi^2$ when comparing the mean clustering of the mock samples to the observed is close to 1 per degree of freedom for $0 < k < 0.3 h$Mpc$^{-1}$ ($\chi^2$/dof = 33/32) and $0 < s < 200h^{-1}$Mpc ($\chi^2$/dof = 18/24). This implies that these mock galaxy samples are appropriate for determining our covariance matrices and testing our methodology.

\subsection{Testing BAO measurements on mocks}
We test our methodology and characterise our expected results by measuring the BAO scale of the mock galaxy samples. We quote results as the best-fit value with 1 and 2 $\sigma$ uncertainty defined as half of the width of the $\Delta\chi^2 = 1$ and 4 regions. First, we fit the mean of the mock samples, using the covariance matrix expected for one realisation. Pre-reconstruction, we find $\alpha =0.998\pm0.080$ for $P(k)$ (no lower 2$\sigma$ bound, upper at 1.19) and $\alpha = 1.005\pm0.095$ for $\xi(s)$ (no upper 2$\sigma$ bound, lower at 0.81). Accurate BAO measurements are therefore not expected to be achievable with the pre-reconstruction data. Post-reconstruction, substantial improvement is observed. For $\xi(s)$, we find $\alpha = 0.998\pm0.048(\pm0.116)$ for the 1 (2)$\sigma$ uncertainty and for $P(k)$ we find $\alpha = 0.998\pm0.044(\pm0.103)$. A decrease in the recovered BAO scale post-reconstruction is expected, as reconstruction removes a small shift in the BAO scale due to mode-coupling (see, e.g., \citealt{CS06,Eis07rec,Pad09,Mehta11}). We focus solely on the results found using the post-reconstruction data in what follows.

\begin{table}
\caption{The statistics of BAO scale measurements recovered from the reconstructed mock samples, excluding 4$\sigma$ outliers and realisations with low BAO detection significance. $S_\alpha$ is the standard deviation of $\alpha$ values between the mocks, the $\langle \sigma \rangle$ is the mean difference in $\alpha$ with $\Delta\chi^2=1$,  $\langle 2\sigma \rangle$ is the mean difference in $\alpha$ with $\Delta\chi^2=4$ and $N$ is number of mocks after excluding outliers and low BAO detection cases. }
\begin{tabular}{lcccccccc}
\hline
\hline
Case  &  $\langle \alpha \rangle$ & $S_{\alpha}$ &  $\langle \sigma \rangle$ &  $\langle 2\sigma \rangle$ & $\langle \chi^2 \rangle$/dof  & $N$  \\
\hline
$P(k)+\xi(s)$ & 0.996 & 0.047 & 0.042 & 0.092 & - & 895 \\
combined $P(k)$ & 0.996 & 0.046  & 0.042 & 0.093 & - & 894 \\
combined $\xi(s)$ & 0.997 & 0.046 & 0.042 & 0.092 & - & 879\\
$P(k)$ & 0.997 & 0.045  & 0.042 & 0.091 & 26.3/27 & 879 \\
$\xi(s)$ & 0.997 & 0.047  & 0.041 & 0.091 & 15.9/16 & 867 \\
\hline
\label{tab:mockbao}
\end{tabular}
\end{table}

Statistics describing the BAO measurements determined from reconstructed mock realisations are presented in Table \ref{tab:mockbao}. The BAO detection is conventionally quoted as a $\chi^2$ difference between the best-fit model and the model with no BAO peak. Some of our mocks result in a very low detection of the BAO. A small number of our mocks have a reasonable BAO detection but result in $\alpha$ values that are far from the mean. These extreme cases tend to correspond to low values of $\alpha$ and artificially skew the recovered distribution. To avoid this problem we only quote the numbers for mocks that have a 2$\sigma$ bound and exclude $>4\sigma$ outliers. The values of the mean recovered $\alpha$, the recovered uncertainties, and the standard deviations are all consistent to within 5 per cent for the fiducial $P(k)$ and $\xi(s)$ measurements, suggesting our method of determining the BAO scale works equally well in configuration and Fourier space.

As in \cite{alphdr11}, we combine results across bin centres. We consider four bin centres for $\xi(s)$ (separated by 2$h^{-1}$Mpc) and 5 for $P(k)$ (separated by 0.0016$h$Mpc$^{-1}$). For $\xi(s)$, the correlation between the results is between 0.91 and 0.95. For $P(k)$, very little variation is expected with bin centre, as the correlations are between 0.98 and 0.99. To combine the results across the bins centres, for both $P(k)$ and $\xi(s)$ we take the mean of the $\Delta\chi^2$ distributions to obtain an averaged $\Delta\chi^2$ distribution. These results are denoted in Table \ref{tab:mockbao} as `combined'. Due to the large correlation between these measurements, there is no significant improvement in the standard deviation, for either $P(k)$ or $\xi(s)$. However, the number of outliers and non-detections decreases, by 15 for $P(k)$ and 12 for $\xi(s)$. Finally, the correlation between the combined $P(k)$ and $\xi(s)$ BAO measurements recovered from the 1000 mocks is 0.97. Given such a large correlation, we take the mean of the combined $\xi(s)$ and $P(k)$ $\Delta\chi^2$ to obtain our $P(k)+\xi(s)$ results. 

After combining measurements to obtain the $P(k)+\xi(s)$ results, we still have 105 cases that either have no 2$\sigma$ bound or are $>4\sigma$ outliers. This suggests the results we quote in Table \ref{tab:mockbao} are slightly optimistic. Considering the full set of 1000 combined mock results, we find that the 68 percentile bounds of the maximum-likelihood $\alpha$ distribution are (0.948,1.040) and that the mean 1$\sigma$ uncertainty is 0.045. The half-width of the 68 percentile region and the mean 1$\sigma$ uncertainty for the full set are thus both close to the standard deviation of $\alpha$ values we report in Table \ref{tab:mockbao}, suggesting the quoted results are a fair representation of the typical likelihood distribution we expect to recover.

As expected based on the fits to the mean $\xi(s)$ and $P(k)$, far more results are found at the tails of the distribution than would be expected if they followed a Gaussian distribution. The individual likelihood distributions are thus not-well represented by a Gaussian either; the mean 2$\sigma$ interval ($\Delta\chi^2 = 4$) is greater than twice the mean $1\sigma$ interval ($\Delta\chi^2=1$). Further, the standard deviation of the best-fit $\alpha$ is closer to half of the mean 2$\sigma$ width than the 1$\sigma$. Thus, in order to use these BAO measurements to constrain cosmological parameters, the full likelihood distribution should be used, rather than a Gaussian approximation. 

The mean $\alpha$ we recover from our mock samples for $P(k)+\xi(s)$ is more than 2$\sigma$ away (1$\sigma$ calculated via $0.047/\sqrt{895}$) from 1. This result is partially driven by outliers as, if we exclude fifteen additional $> 3\sigma$ outliers, the mean increases to 0.997 (and the standard deviation decreases to 0.044). However, the small bias remains. The magnitude of this bias is less than 0.1 of the 1$\sigma$ expected uncertainty in the measurement from the data, and thus not significant in any application of our measurements. Given that we use the same methodology as \cite{alphdr11} who found no detectable bias in isotropic BAO measurements, we do not believe this bias is suggestive of a systematic bias in the BAO fitting methodology.

\section{Results}
\label{sec:results}
\subsection{Distance Scale Measurement}

\begin{figure}
\includegraphics[width=84mm]{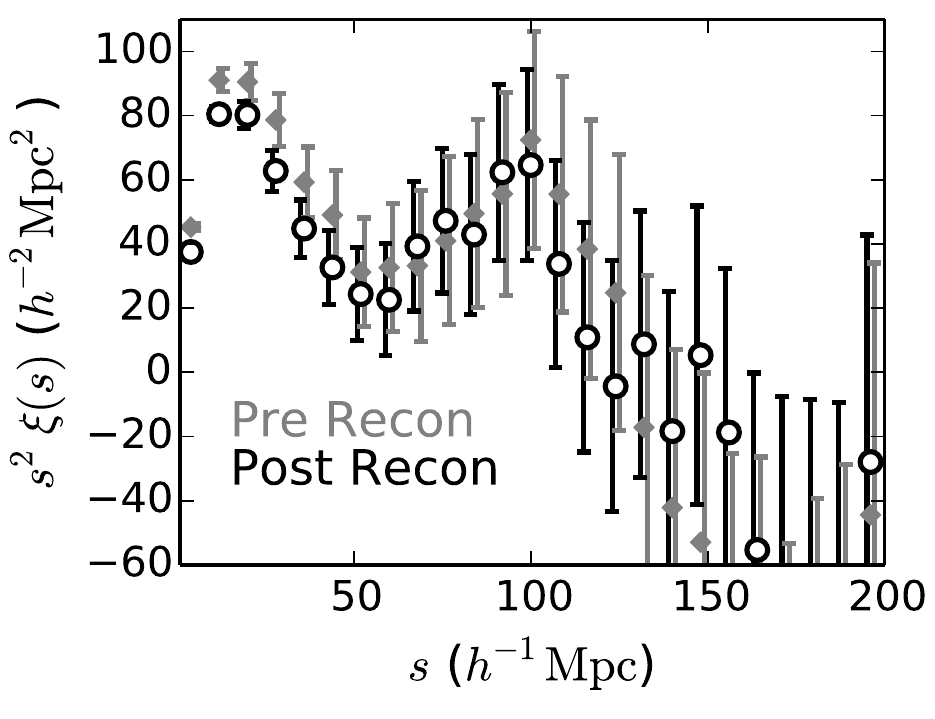}
  \caption{The measured correlation function, $\xi(s)$ (points with error-bars), pre- (grey diamonds) and post- (open circles) reconstruction. The error-bars are determined from the variance of the 1000 mock galaxy samples. One can see that reconstruction reduces the clustering amplitude, due to removal of large-scale redshift space distortions, and sharpens the BAO peak.}
\label{fig:comxi}
\end{figure}

\begin{figure}
\includegraphics[width=84mm]{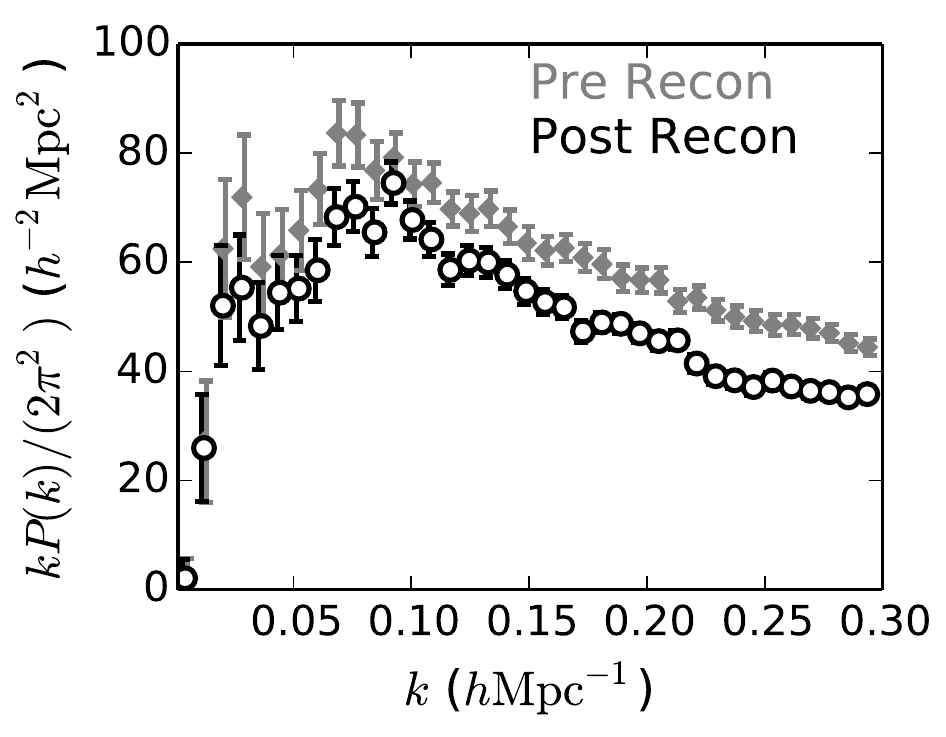}
  \caption{The measured power spectrum, $P(k)$ (points with error-bars), pre- (grey diamonds) and post- (black circles) reconstruction. The error-bars are determined from the variance of the 1000 mock galaxy samples. The clustering amplitude of the post-reconstruction data is decreased across all $k$ due to the removal of large-scale redshift space distortions.}
\label{fig:compk}
\end{figure}

The clustering of our galaxy sample, pre-(grey diamonds) and post-reconstruction (open circles), is displayed in Fig. \ref{fig:comxi} for the correlation function and in Fig. \ref{fig:compk} for the power spectrum. One can see, most easily by studying the $P(k)$ measurements, that reconstruction induces a decrease in the clustering amplitude that is nearly fractionally constant (at scales less than the BAO scale for $\xi[s]$). This is due to the removal of large-scale redshift-space distortions. One can further see, most easily by studying the $\xi(s)$ measurements, that reconstruction sharpens the BAO feature. Throughout the rest of the section, we focus our attention on BAO measurements obtained from the post-reconstruction measurements.

\begin{figure}
\includegraphics[width=84mm]{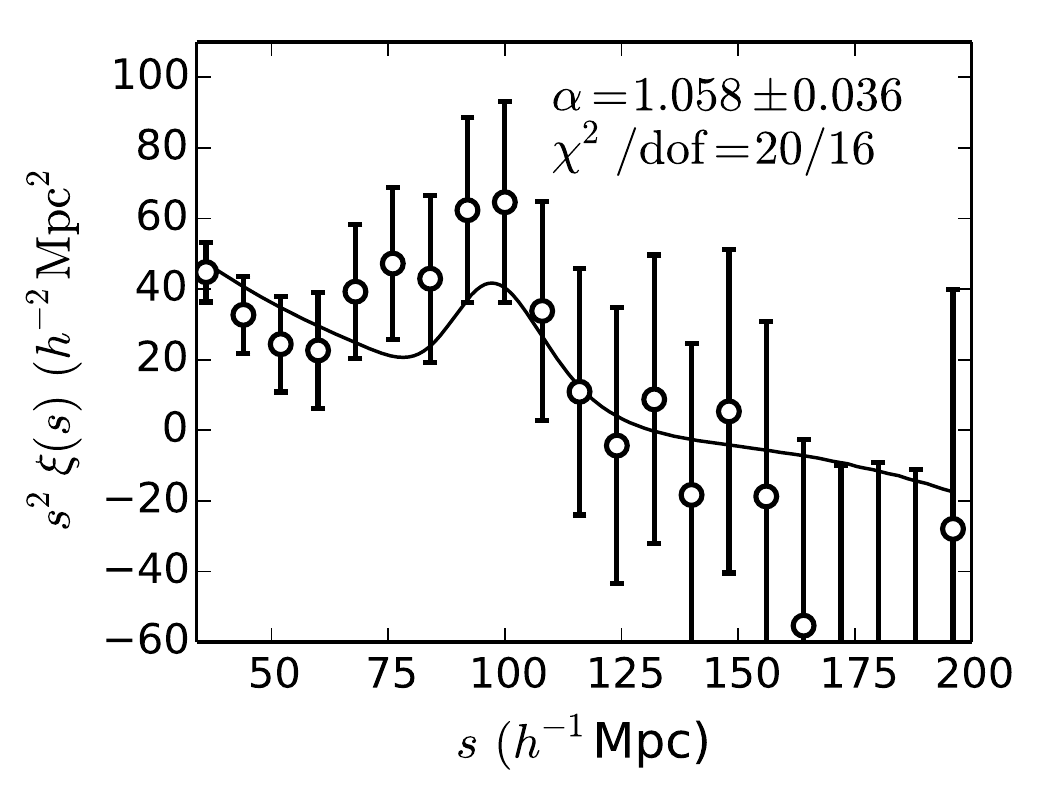}
  \caption{The measured post-reconstruction correlation function, $\xi(s)$, (points with error-bars) and best-fit BAO model (solid curve) . }
\label{fig:xibao}
\end{figure}

\begin{figure}
\includegraphics[width=84mm]{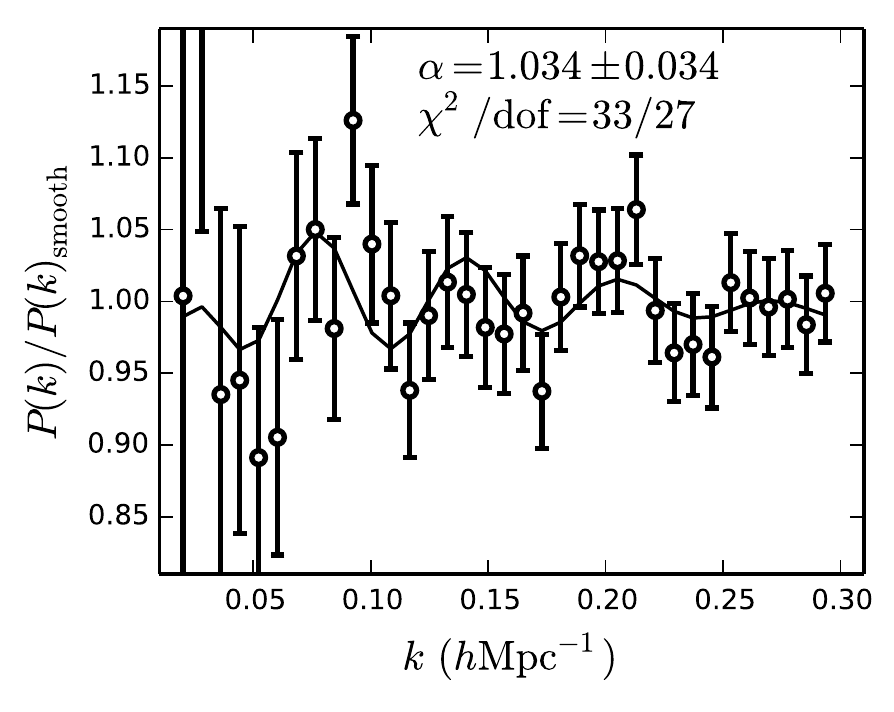}
  \caption{The measured post-reconstruction power spectrum, $P(k)$, (points with error-bars) and best-fit model (solid curve) divided by the smooth (no BAO) component of the best-fit model. }
\label{fig:Pkbao}
\end{figure}

We obtain detections of the BAO signal for our reconstructed data using both $P(k)$ and $\xi(s)$ and we combine these data to obtain our consensus measurement of $\alpha = 1.040\pm0.037$. The best-fit model found for the fiducial $\xi(s)$ binning is displayed against the measurement in Fig. \ref{fig:xibao}. We find $\alpha = 1.058\pm0.036$. The $\chi^2$/dof is slightly greater than one (20.3/16) and represents a good fit (a greater $\chi^2$ would be expected in 21 per cent of cases). The best-fit $P(k)$ BAO model (divided by the smooth component of the best-fit $P(k)$ model), using our fiducial binning, is displayed in Fig. \ref{fig:Pkbao}. We find $\alpha = 1.034\pm0.034$ for $P(k)$. Similar to $P(k)$, the best-fit model is a good fit ($\chi^2$/dof = 32.6/27, a greater $\chi^2$ would be expected in 21 per cent of cases). The measured BAO positions are consistent with those observed in the pre-reconstruction clustering measurements, listed in Table \ref{tab:baoresults}, but reconstruction improves the uncertainty by greater than a factor of 2 (2.4 for $\xi(s)$ and 2.7 for $P(k)$). 
The improvement is due to the sharpening of the BAO feature post-reconstruction as can be observed in Fig.~\ref{fig:comxi}: in turn we fit this with a template with a lower value of $\Sigma_{nl}$ (matching the best-fit value found for the mock samples). The improvement is greater than the mean factor of $\sim$ 2 expected based on the mean of the mock samples, but not unprecedented; many realizations are found to improve by greater than a factor of 3 in \cite{alph}, though the mean improvement is by a factor of 1.5.
\begin{table}
\centering
\caption{Isotropic BAO scale measurements. The `Consensus' results are the mean of the combined $P(k)$ and $\xi(s)$ results. The uncertainties quoted are the 1$\sigma$(2$\sigma$) intervals determined via $\Delta\chi^2=1(4)$. The $P(k)$ measurements are numbered based on the shift in the bin centre, which are integer multiples of $0.0016h$Mpc$^{-1}$. For $\xi(s)$, the results are numbered based on the bin centre shift in $h^{-1}$ Mpc. The measurements using each bin centre are averaged, as described in the text, in order to obtain the `Combined' measurements.}
\begin{tabular}{llc}
\hline
\hline
Estimator  &   $\alpha$ & $\chi^2$/dof \\
\hline
{\bf Consensus} & {\bf $1.040\pm0.037(0.084)$} & \\
Combined $P(k)$ & $1.031\pm0.034(0.077)$\\
Combined $\xi(s)$ & $1.050\pm0.040(0.092)$\\
post-recon 0 $P(k)$ & $1.034\pm0.034$ & 32.6/27  \\
post-recon 1 $P(k)$ & $1.030\pm0.034$ & 26.6/27 \\
post-recon 2 $P(k)$ & $1.027\pm0.034$ & 23.9/27 \\
post-recon 3 $P(k)$ & $1.030\pm0.034$ & 27.4/27 \\
post-recon 4 $P(k)$ & $1.033\pm0.034$ & 30.9/27 \\
post-recon 0 $\xi(s)$ & $1.058\pm0.036$ &  20.3/16 \\
post-recon 2 $\xi(s)$ & $1.045\pm0.037$ &  21.0/17 \\
post-recon 4 $\xi(s)$ & $1.040\pm0.044$ &  25.3/16 \\
post-recon 6 $\xi(s)$ & $1.056\pm0.043$ &  21.1/16 \\
pre-recon $P(k)$ & $1.033\pm0.093$ & 19.1/27   \\
pre-recon $\xi(s)$ & $1.013\pm0.094$ & 12.2/17 \\ 
\hline
\label{tab:baoresults}
\end{tabular}
\end{table}

Post-reconstruction BAO measurements for each of the bin centres we consider are listed in Table \ref{tab:baoresults}. More variation is observed for $\xi(s)$ than for $P(k)$, which is expected, as the measurements using different $\xi(s)$ bin centres were found to be less correlated in the mock samples than the $P(k)$ measurements. The differences we find in the $\xi(s)$ BAO measurements are as large as 0.018. This is consistent with the correlation factors that we find in the mock samples (between 0.91 and 0.95), which suggest the 1$\sigma$ scatter between bin centre results varies between 0.013 and 0.017 for a 0.04 statistical uncertainty.

\begin{figure}
\includegraphics[width=84mm]{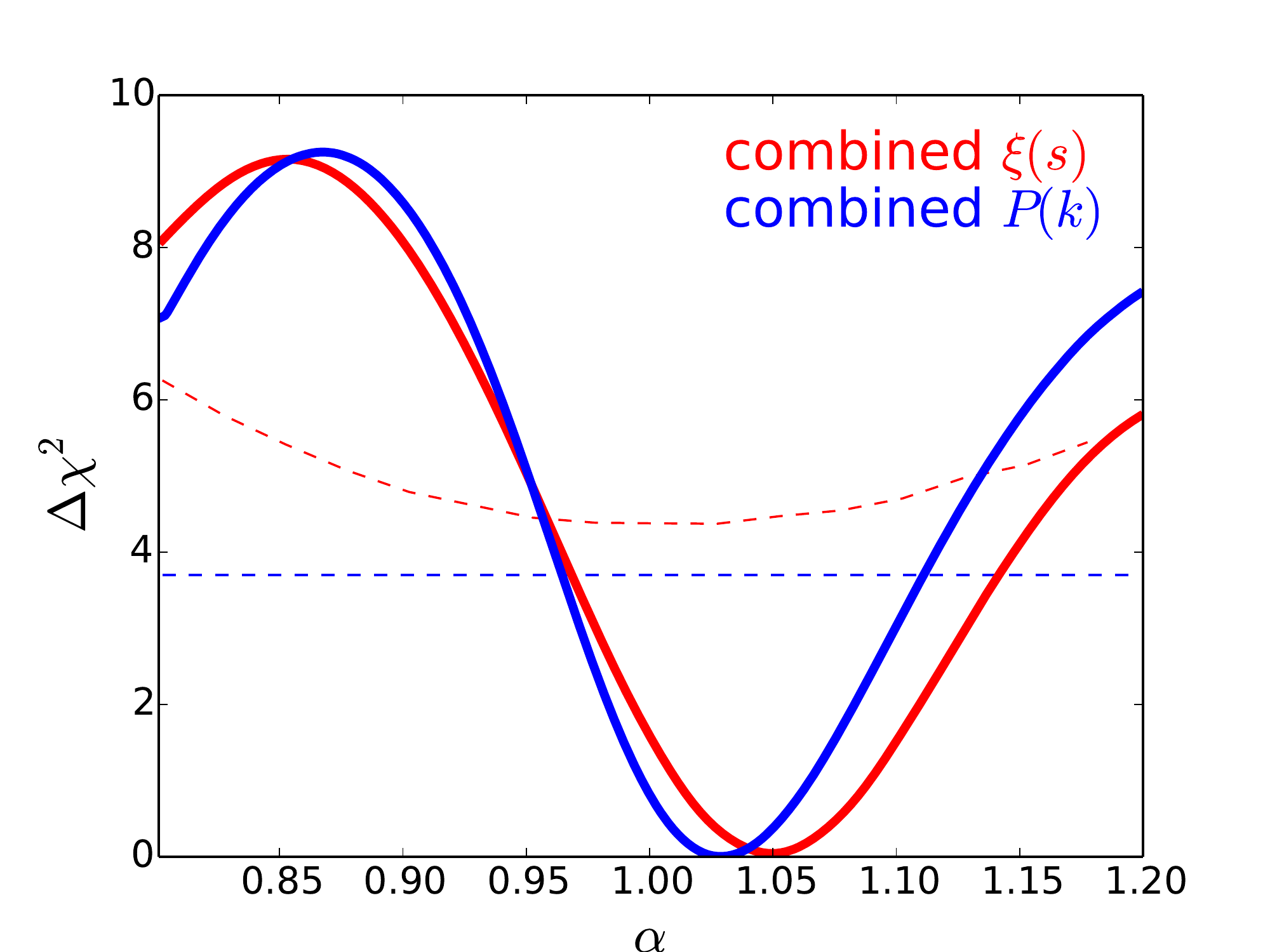}
  \caption{The solid curves show $\Delta\chi^2$ as a function of the BAO scale $\alpha$ for the combined $P(k)$ and $\xi(s)$ measurements. The dashed lines show the $\Delta\chi^2$ between the best-fit BAO model and the model without any BAO, using the fiducial bin choice. The $\xi(s)$ and $P(k)$ measurements are consistent with each other, and both prefer the BAO model at close to 2$\sigma$ ($\Delta\chi^2=4$) significance.}
\label{fig:BAOdet}
\end{figure}

For both $P(k)$ and $\xi(s)$, the post-reconstruction measurements are combined by taking the mean of the $\Delta\chi^2$ across the bin centres listed in Table \ref{tab:baoresults}. These averaged results are listed as `Combined' in Table \ref{tab:baoresults}. We display the $\Delta\chi^2$ for each in Fig. \ref{fig:BAOdet}, using solid curves. One can see that the two likelihoods are consistent and that the $P(k)$ measurement is slightly more precise. The dashed lines display the $\Delta\chi^2$ between the best-fit BAO model and the model without any BAO, using the fiducial bin choice. Both $\xi(s)$ and $P(k)$ prefer the BAO model at close to 2$\sigma$. The preference is less significant for $P(k)$. This is due to the fact that the smooth component of the $P(k)$ has five free terms, while for $\xi(s)$, there are only three terms. For the same reason, similar differences were found in \cite{alphdr11}. Further, the smooth $\xi(s)$ model depends on $\alpha$, while the smooth $P(k)$ model does not (see Eqs. \ref{eq:mod_pk} and \ref{eq:pksm}), explaining the shape of each dashed curve.

The $\xi(s)$ ($1.050\pm0.040$) measurement is less precise than its $P(k)$ counterpart ($1.031\pm0.034$). The difference in the uncertainties is typical of what we find in the mocks samples, as we find 28 per cent of the mock realisations have $\sigma_{\xi}/\sigma_P$ or $\sigma_P/\sigma_{\xi}$ that is larger than the $\sigma_{\xi}/\sigma_P$ we find for the data. The difference in the $\alpha$ values of 0.019 is more unusual. The standard deviation between the combined $P(k)$ and $\xi(s)$ found in the mocks is 0.011, implying the difference is $\sim 2\sigma$. Counting the number of mocks that have a larger value of $(\alpha_{\xi}-\alpha_P)^2/(\sigma^2_\xi+\sigma^2_P)$, we find 3 per cent of the realisations have a larger difference. The difference between the $P(k)$ and $\xi(s)$ is of similar significance as was found in \cite{alphdr11}. 

The differences between $P(k)$ and $\xi(s)$ BAO measurements are unusually large for both our data and those of \cite{alphdr11} compared to the differences found in the mocks due to the large correlations (0.97 for our mocks and 0.95 for the \citealt{alphdr11} mocks) achieved in the mock samples. The precision of our measurements is a factor of four less than that of \cite{alphdr11}, suggesting that a common explanation requires greater scatter between the results obtained from mock samples (rather than a bias in one particular measurement technique). For our MGS data, the difference we find would be 1$\sigma$ if the correlation were reduced to 0.88. It is conceivable that including more realism (e.g., light-cones) in mock samples would reduce the correlation between the $P(k)$ and $\xi(s)$ BAO measurements recovered from such mocks, which would explain both our findings and those of \cite{alphdr11}. The robustness checks we describe in the following section reveal no potential systematic effect that would bias either the $\xi(s)$ and $P(k)$ measurements. We therefore obtain our consensus result of $\alpha = 1.041\pm0.037$ by taking the mean of the combined $P(k)$ and $\xi(s)$ $\Delta\chi^2$.

The consensus likelihood we obtain is non-Gaussian. Therefore, one should use the full likelihood (and not a Gaussian approximation) when testing any cosmological model. A table containing $\alpha$, as defined by Eq. \ref{eq:alpha} and gridded with spacing 0.001, is provided as a supplementary file in the online publication\footnote{It is also included in the source files of the arXiv submission, named `chid\_MGSconsensus.dat'}. The first five rows of this table are shown in Table \ref{tab:alphachi2}.
\begin{table}
\centering
\caption{The first five rows (the complete version is available as a supplementary file in the online publication) of a table containing our consensus $\Delta\chi^2$ at each $\alpha$, defined by Eq. \ref{eq:alpha}, gridded with a spacing 0.001. For our fiducial cosmology, $D^{\rm fid}_V(z) = 638.95$ Mpc and $r_d^{\rm fid} = 148.69$ Mpc. Given that our consensus likelihood is not well represented as a Gaussian, this table should be used to test any cosmological model.}
\begin{tabular}{cc}
\hline
\hline
   $\alpha$ & $\Delta\chi^2$ \\
\hline
0.8015 & 7.80363740636 \\
0.8025 & 7.7078667921 \\
0.8035 & 7.74064378608 \\
0.8045 & 7.78165406477 \\
0.8055 & 7.8259614948 \\
\hline
\label{tab:alphachi2}
\end{tabular}
\end{table}

\subsection{Robustness Tests}
We perform a series of robustness tests on BAO scale measured using the post-reconstruction $\xi(s)$ and $P(k)$. We test the range of scales used in the fit, the number of terms used in the broadband polynomial, and the damping assumed in the BAO template. The results are shown in Table \ref{tab:robbao}. The $\alpha$ measurements are insensitive to changes in the range of scales that is fit, as we have tested fit ranges of $50 < s < 150h^{-1}$ and $0.05 < k < 0.28h$Mpc$^{-1}$ and have found negligible changes in the best-fit $\alpha$ and its uncertainty.

The measurements of $\alpha$ are robust to the form of the polynomial. For $P(k)$, the data prefer a 5 term polynomial to a three term one, as the $\chi^2$ increases by 7 while adding only two degrees of freedom. These results agree with \cite{alphdr11}, who found that the 5-term polynomial was required to achieve a good-fit both for the mock and data $P(k)$. The best-fit $\alpha$ shifts by 0.1$\sigma$ and the uncertainty increases by 6 per cent when the 3-term polynomial is used, suggesting that while the data prefer the 5-term polynomial, the BAO scale measurements are insensitive to the exact form of the polynomial applied to the $P(k)$ fits. 

The $\alpha$ measurements obtained from $\xi(s)$ are insensitive to whether any smooth polynomial is included in the fit or not.
The $\chi^2$/dof is smaller for $\xi(s)$ when no polynomial is used compared to the fiducial three term polynomial. Adding two additional terms to the polynomial reduces the $\chi^2$ such that the $\chi^2$/dof is the same with and without any polynomial. The best-fit $\alpha$ values change by 0.001 and the uncertainty by only 8 per cent when the form of the polynomial is altered for $\xi(s)$, implying its exact form has negligible influence on our results.

The choices we adopt for the BAO damping parameters have a larger impact on our measurements, especially for the uncertainty. The data has a small preference for a low value of $\Sigma_{nl}$. When allowed to vary freely, the best-fit for both $P(k)$ and $\xi(s)$ is found when $\Sigma_{nl} = 0$. The BAO feature is enhanced in the BAO template when $\Sigma_{nl}$ is lower and thus the uncertainty drops by a factor of 30 per cent. Allowing $B$ to vary freely in the $\xi(s)$ fit has the same effect, as the best-fit increases $B$,  compensating with the polynomial, and thereby increases the size of the BAO feature. However, this lower value of $\Sigma_{nl}$ (or greater value of $B$) is preferred only weakly: the $\chi^2$ decreases by only 1.1 for $\xi(s)$ and 0.7 for $P(k)$. While the uncertainty changes significantly, the best-fit values of $\alpha$ shift by only 0.26$\sigma$ for $P(k)$ and 0.08$\sigma$ for $\xi(s)$ when $\Sigma_{nl}$ is allowed to be free. The damping of the BAO feature is predicted by perturbation theory (see e.g., \citealt{CS06,EisSeoWhi07,Pad09}) and is expected based on both our mock samples and more detailed simulations (e.g., \citealt{Angulo08,Mehta11}). We therefore use the best-fit $\Sigma_{nl}$ recovered from the mean of our mock samples, which we believe is more physically appropriate. Our tests show that the BAO distance scale we measure is robust to this choice, but that allowing no damping would cause us to under-estimate our uncertainty.

\begin{table}
\centering
\caption{Robustness tests performed on the $P(k)$ and $\xi(s)$ BAO measurements recovered from the post-reconstruction data. The fiducial $P(k)$ results are fit in the range $0.02 < k < 0.3 h$Mpc$^{-1}$ using a five-term polynomial and $\Sigma_{nl} = 5.3\pm2h^{-1}$Mpc. The fiducial $\xi(s)$ results are fit in the range $30 < s < 200 h^{-1}$Mpc using a three-term polynomial, $\Sigma_{nl} = 4.5h^{-1}$Mpc, and ${\rm log}(B) = 1\pm0.4$. (All $\pm$ represent Gaussian priors.) }
\begin{tabular}{lccc}
\hline
\hline
Estimator  &   $\alpha$ & $\chi^2$/dof \\
\hline
$P(k)$: & &\\
fiducial & $1.034\pm0.034$ & 32.6/27  \\
$0.05 < k < 0.28h$Mpc$^{-1}$ & $1.033\pm0.032$ & 27.6/21 \\
$A_1,A_2 = 0$ & $1.038\pm0.036$ & 39.7/29 \\
$\Sigma_{nl} = 4.3\pm2h^{-1}$Mpc & $1.030\pm0.031$ & 32.3/27\\
$\Sigma_{nl} = 6.3\pm2h^{-1}$Mpc & $1.038\pm0.038$ & 32.9/27\\
$\Sigma_{nl} = 5.3\pm0h^{-1}$Mpc & $1.035\pm0.035$ & 32.7/28\\
$\Sigma_{nl}$ free & $1.025\pm0.026$ & 31.9/27\\
$\xi_0(s)$: & & \\
fiducial & $1.058\pm0.036$ &  20.3/16 \\
$50 < s < 150h^{-1}$Mpc  & $1.057\pm0.037$ &  13.1/8 \\
$a_1,a_2,a_3 = 0$  & $1.059\pm0.037$ &  20.7/19 \\
$+a_4,a_5$ & $1.059\pm0.039$ & 15.1/14\\
$B$ free & $1.058\pm0.029$ & 19.2/16\\
$\Sigma_{nl} = 3.5h^{-1}$Mpc & $1.056\pm0.035$ & 20.2/16\\
$\Sigma_{nl} = 5.5h^{-1}$Mpc & $1.060\pm0.041$ & 20.7/16\\
$\Sigma_{nl} = 0h^{-1}$Mpc & $1.055\pm0.029$ & 19.2/16\\
\hline
\label{tab:robbao}
\end{tabular}
\end{table}

\section{Cosmological Interpretation}  \label{sec:cosmology}

\subsection{BAO Distance Ladder}
\begin{figure}
\includegraphics[width=84mm]{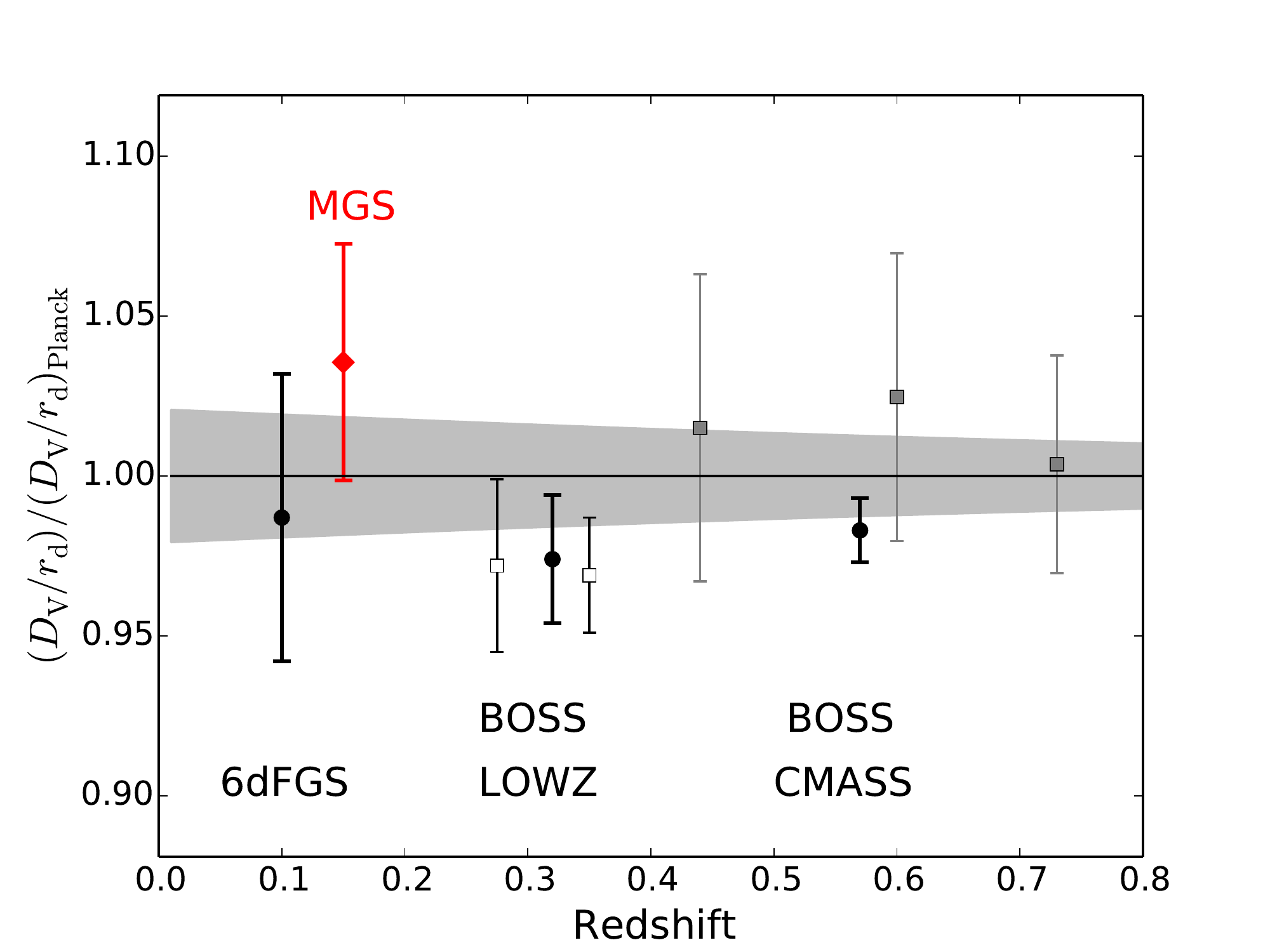}
  \caption{The BAO distance ladder, expressed as $D_V/r_d$, including our measurement and relative to the Planck prediction given their best-fit flat $\Lambda$CDM model. The grey region represents the 1$\sigma$ uncertainty given Planck data and assuming a flat $\Lambda$CDM model. Our measurement, using the SDSS DR7 main galaxy sample, is displayed with a red diamond. Measurements made using 6dFGS data \citep{Beutler11} and BOSS data (\citealt{alphdr11,Toj14}) are denoted with black circles. These measurements are nearly independent with ours, allowing them to be combined to obtain cosmological constraints. The white squares display measurements using SDSS DR7 data (\citealt{Per10,Xu12}) and the grey squares display measurements made using WiggleZ data \citep{Kazin14}.}
\label{fig:BAOlad}
\end{figure}

Our measurement provides a new rung for the BAO distance ladder. Fig. \ref{fig:BAOlad} shows current BAO-scale measurements compared with our MGS measurement, displayed using a red diamond. The measurements in Fig. \ref{fig:BAOlad} are divided by the prediction for the best-fit flat $\Lambda$CDM model, as determined from Planck satellite (\citealt{PlanckOver,PlanckCos}) observations of the CMB. The grey contour represents the 1$\sigma$ allowed region, determined by sampling the Planck likelihood chains.\footnote{We used Planck $\Lambda$CDM chains base-planck-lowl-lowLike-highL which are publicly available for download from Plank Legacy Archive at http://pla.esac.esa.int/pla/aio/planckProducts.html at the moment of writing this paper.} The points displayed using black circles form a set of independent measurements; these include the 6dFGS measurement made by \cite{Beutler11} and BOSS measurements made by \cite{Toj14} and \cite{alphdr11}. We combine these data with our own to obtain cosmological constraints in the following section. The white squares represent measurements made using SDSS DR7 data (\citealt{Per10} and \citealt{Xu12}) and the grey squares are measurements made by \cite{Kazin14} using WiggleZ data. These data overlap significantly in volume with the BOSS data and we do not use them to obtain cosmological constraints. The BAO distance measurements are broadly consistent with each other and the Planck best-fit $\Lambda$CDM prediction.

The volume of our MGS sample overlaps slightly with 6dFGS. The 6dFGS footprint occupies most of the Southern sky, i.e., it has $\delta < 0$. Less than 3 per cent of our sample has $\delta < 0$. This, combined with the fact that the redshift distributions of the 6dFGS data and our MGS data are significantly different, suggests that any covariance in our BAO measurement with the \cite{Beutler11} measurement is negligible.

The footprint of our sample has a large overlap with the footprint of the BOSS LOWZ sample used in \cite{Toj14}. It also overlaps in redshift in the range $0.15 < z < 0.2$. The effective redshift of the LOWZ measurement is $z_{eff} = 0.32$ and the data with $0.15 < z < 0.2$ is only a small fraction of the total volume covered by LOWZ. Calculating the effective volume of LOWZ sample using \citep{Tegmark98}
\begin{equation}
v_{eff} = \int \left[\frac{n(r)P_0}{1+n(r)P_0}\right]^2d^3r,
\end{equation}
and $P_0 = 20,000h^3$Mpc$^{-3}$, we find 8 per cent of the LOWZ volume is at $0.15 < z < 0.2$. Accounting for the fact that more than one quarter of the BOSS LOWZ data is in the SGC and that our footprint does not perfectly overlap with the BOSS LOWZ NGC footprint, the volume of the BOSS LOWZ sample at $0.15 < z < 0.2$ and overlapping with our footprint is less than 5 per cent. Using $P_0 = 16000h^3$Mpc$^{-3}$, 58 per cent of the volume of our sample is at $0.15 < z < 0.2$. This implies the total volume overlap between our sample and the BOSS LOWZ sample used in \cite{Toj14} is 3 per cent and thus the correlation between the respective BAO measurements is less than 0.03 (it will be smaller than the volume overlap due to the difference in the galaxy samples used). This is small enough to be negligible when determining cosmological constraints. 

\subsection{Cosmological Constraints with BAO}
\begin{figure}
\includegraphics[width=84mm]{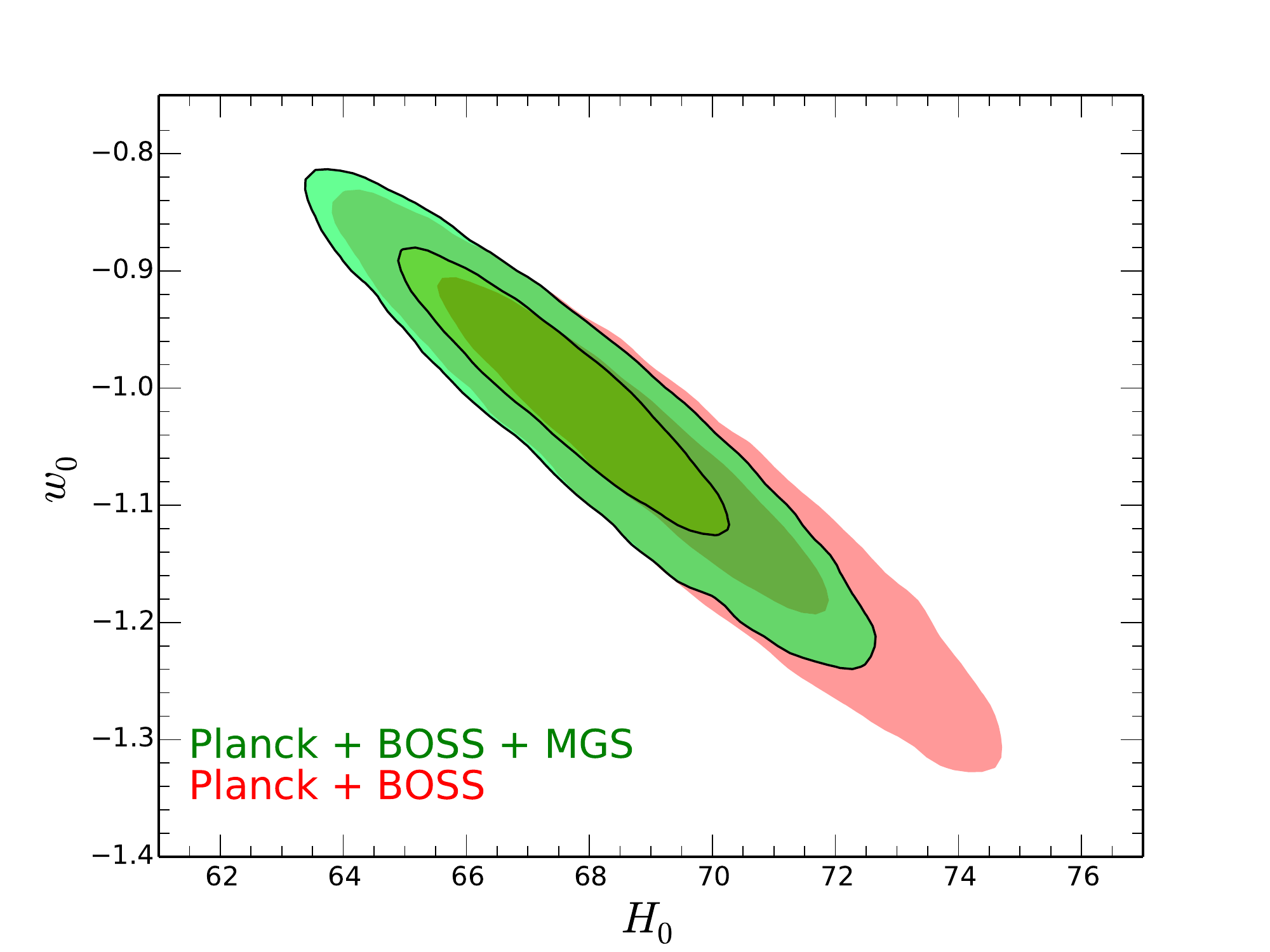}
  \caption{The 1 and 2$\sigma$ confidence levels for the dark energy equation of state, $w_0$, and the value of the Hubble constant, $H_0$, constraints combining BAO distance measurements with Planck data. We show the results when including Planck and BOSS data (red) and then when also including our measurement made using SDSS DR7 MGS data (green). The inclusion of our measurement decreases the area enclosed by the 1$\sigma$ contour by 20 per cent. }
\label{fig:wOm}
\end{figure}

\begin{table*}
\centering
\begin{tabular}{lllllll}
\hline
Cosmological & Data Sets & $\Omega_{\rm m} h^{2}$ & $\Omega_{\rm m}$ & $H_{0}$ & $\Omega_{\rm K}$ & $w_{0}$ \\
Model & & & & km s$^{-1}$ Mpc$^{-1}$ & &  \\
\hline
$\Lambda$CDM & Planck & 0.1427 (26) & 0.316 (16) & 67.3 (12) & - & - \\
\hline
$\Lambda$CDM & Planck + BOSS  & 0.1414 (14) & 0.308 (9) & 67.7 (6) & - & -  \\
$\Lambda$CDM & Planck + {\bf MGS} & 0.1435 (23) & 0.323 (15) & 66.7 (11) & - & - \\ 
$\Lambda$CDM & Planck + BOSS + {\bf MGS} & 0.1418 (14) & 0.311 (8) & 67.6 (6) & - & - \\
$\Lambda$CDM & Planck + BOSS + 6dF + {\bf MGS} & 0.1418 (13) & 0.311 (8) & 67.6 (6)  & - & - \\
\hline
o$\Lambda$CDM & Planck + BOSS  & 0.1419 (25) & 0.309 (9) & 67.8 (8) & +0.0005 (30) & - \\
o$\Lambda$CDM & Planck + {\bf MGS} & 0.1418 (25) & 0.351 (24) & 63.7 (21) & -0.0095 (58) & - \\ 
o$\Lambda$CDM & Planck + BOSS + {\bf MGS} & 0.1420 (26) & 0.311 (9) & 67.6 (8) & +0.0002 (31) & - \\
o$\Lambda$CDM & Planck + BOSS + 6dF + {\bf MGS}& 0.1421 (26) & 0.311 (8) & 67.6 (7) & +0.0002 (30) & - \\
\hline
$w$CDM & Planck + BOSS  & 0.1425 (22) & 0.300 (17) & 69.1 (22) & - & -1.064 (101) \\
$w$CDM & Planck + {\bf MGS} & 0.1433 (24) & 0.324 (48) & 67.2 (60) & - & -1.006 (195) \\
$w$CDM & Planck + BOSS + {\bf MGS} & 0.1420 (22) & 0.309 (15) & 67.8 (19) & - & -1.013 (86) \\
$w$CDM & Planck + BOSS + 6dF + {\bf MGS}& 0.1420 (21) & 0.310 (14) & 67.7 (17) & - & -1.010 (81) \\
\hline
o$w$CDM & Planck + BOSS & 0.1419 (25) & 0.296 (24) & 69.3 (28) & -0.001 (4) & -1.08 (15) \\
o$w$CDM & Planck + BOSS + {\bf MGS} & 0.1420 (25) & 0.312 (20) & 67.5 (22) & +0.001 (5) & -1.00 (13) \\
o$w$CDM & Planck + BOSS + 6dF + {\bf MGS} & 0.1421 (25) & 0.313 (21) & 67.6 (23) & +0.001 (5) & -0.99 (13) \\
\hline
\end{tabular}
\caption{Constraints for cosmological parameters using different combinations of BAO and Planck+WP+highL CMB (denoted Planck) data, for models that assume cold dark matter (CDM) and allow various degrees of freedom in curvature and the dark energy equation of state. $\Lambda$CDM assumes a flat geometry ($\Omega_k = 0$) and a cosmological constant ($w=-1$). Models denoted with an `o' allow free $\Omega_k$. Models with `w' allow freedom the equation of state of dark energy. `BOSS' denotes that we use the \citep{Toj14} and \citep{alphdr11} anisotropic BAO measurements. `6dF' denotes that we use the 6dFGS BAO measurement (Beutler et al. 2011). Our measurement made using the SDSS DR7 main galaxy sample is denoted {\bf MGS}.}
\label{tab:cos}
\end{table*}

We identify four independent galaxy BAO measurements that can be combined to obtain cosmological constraints. These include three spherically-averaged measurements; our own MGS measurement at $z=0.15$, \cite{Beutler11} at $z=0.11$, \cite{Toj14} at $z=0.32$, and the anisotropic measurement of \cite{alphdr11} at $z=0.57$.  We combine these data with the CMB results released by \cite{PlanckCos} that are based on the combination of data from the Planck Satellite, Wilkinson Microwave Anisotropy Probe (WMAP) satellite \citep{Bennett03,Spergel03} polarization measurements \citep{Bennett12}, and high-$\ell$ power spectra data from ACT \citep{Das13} and SPT \citep{Story12} and denoted `Planck+WP+highL' in \cite{PlanckCos}. We refer to this combination of CMB data simply as `Planck'. We determine likelihoods for cosmological parameters for the BAO$+$ Planck data set using the {\sc cosmomc} software package \citep{cosmomc1,cosmomc2}. A study by the \cite{Joint} explores the constraints that are achieved when also including BOSS Lyman $\alpha$ forest BAO measurements \citep{FR14,Delubac14} and when considering many extensions to the basic $\Lambda$CDM model. Here, we consider only simple extensions of the $\Lambda$CDM cosmological model and use only the combination of galaxy BAO and CMB measurements, allowing us to focus on the improvement our new measurement provides in determining basic dark energy properties.

Table \ref{tab:cos} presents the maximum likelihood and 68 per cent confidence regions we determine for cosmological parameters, using different combinations of the Planck, BOSS, 6dFGS, and our MGS measurement. The MGS measurement of $D_v/r_d$ is greater than predicted by the Planck best-fit $\Lambda$CDM measurement. In the context of $\Lambda$CDM, this implies a greater value of $\Omega_m$ and a lower value of $H_0$ (both with and without curvature), as we find for the cases where we combine Planck$+$MGS data. However, for any $\Lambda$CDM model, our measurement provides only minor (at best) improvement in the constraints over what Planck$+$BOSS achieves. Essentially, the Planck$+$BOSS measurements fix $\Omega_m$ and this allows very little freedom in the distance-redshift relationship (compared to the precision of our measurement) when the equation of state of dark energy is fixed at -1.

 When we allow the equation of state of dark energy to vary, our BAO measurement provides significant improvement in the precision of $\Omega_m$, $H_0$, and $w_0$. Adding our measurement to either the Planck$+$BOSS or Planck$+$BOSS$+$6dF data sets results in a 15 per cent improvement in the precision the $H_0$ and $w_0$ measurements. This is illustrated in Fig. \ref{fig:wOm}, where the 1 and 2$\sigma$ allowed regions for $w_0$ and $H_0$ are displayed for Planck$+$BOSS (red) and Planck$+$BOSS$+$ MGS (green).
 
In all of the cases we compare, $H_0$ decreases when we include our MGS measurement. For example, in the o$w$CDM case, we find $H_0=67.5\pm2.2$ for the combination of Planck$+$BOSS$+$MGS data, while excluding the MGS measurement yields $H_0=69.3\pm2.8$. The MGS BAO measurement therefore increases the tension between Planck$+$BAO measurements of $H_0$, and those obtained using direct detection, e.g., the measurements by \cite{Efst} of $H_0 = 72.5\pm2.5$,  \cite{Riess11} of $H_0 = 73.8\pm2.4$, and \cite{Freed12} of $H_0 = 74.3\pm2.1$. The constraint on $H_0$ obtained using BAO measurements (including our own) is explored in much greater detail in a study by the \cite{Joint}.  

Using the full data set, we find $w_0=-1.010\pm0.081$ and the best-fit cosmological parameters differ from the the Planck $\Lambda$CDM best-fit by less than 0.4$\sigma$ (where $\sigma$ is the uncertainty on the Planck best-fit measurements). Our measurement thus affords significant improvement in measurements of the properties of dark energy and, in combination with other BAO data, is in excellent agreement with a flat $\Lambda$CDM model. See \cite{Joint} for further exploration of the cosmological implications of BAO measurements, including our MGS measurement.

\section{Conclusion}

We have reanalysed the clustering of galaxies from the SDSS DR7 main galaxy sample (MGS) using state of the art techniques to reconstruct the density field and determine errors using a suite of 1000 mock realisations of our data. Applying these techniques allows a robust, 4 per cent measurement of the BAO scale at $z=0.15$. Our results can be summarised as follows:

$\bullet$ We use the NYU-VAGC `safe0' sample to select a sample of galaxies, with $g-r > 0.8$, $M_r < -21.2$, and $0.07 < z < 0.2$, that occupy a volume not probed by BOSS and have a large enough number density to ensure clustering measurements using the sample are cosmic-variance limited. We denote our sample `MGS'.

$\bullet$ We use 1000 mock realisations of our data, created and validated as described in \cite{Howlett14a,Howlett14b}, in order to produce covariance matrices and test our methodology. Testing the mean clustering of these mock realisations, we expect reconstruction to improve the precision of our BAO measurement by a factor of two. Testing the reconstructed results for individual mocks, we find 90 per cent of the realisations provide robust BAO detection and that measurements determined using Fourier- and configuration-space clustering are highly consistent (0.97 correlation factor).

$\bullet$ We find that reconstruction applied to our data improves the precision of our BAO measurement by greater than a factor of two for both our Fourier- and configuration-space measurements, which we show to be robust against choices in the fitting methodology. These BAO measurements are consistent with each other and we combine them to obtain our consensus measurement of $D_{V}(z_{\rm eff}=0.15) =  (664\pm25)(r_d/r_{d,{\rm fid}})$ Mpc. The likelihood for our measurement is not well approximated as a Gaussian. Instead, one should use the full likelihood, available in the online publication\footnote{It is also included in the source files of the arXiv submission, named `chid\_MGSconsensus.dat'} as detailed in Table \ref{tab:alphachi2}.

$\bullet$ Our distance scale measurement can be combined with Planck CMB data and other BAO distance scale measurements to improve the precision of cosmological constraints. Our distance measurement is larger than that predicted by Planck data and a $\Lambda$CDM model, and therefore decreases the derived value of $H_0$ when combining CMB and BAO data. Thus, our measurement increases the tension between direct $H_0$ measurements and CMB+BAO derived constraints on $H_0$. For dark energy constraints, including our measurement in addition to BOSS and 6dFGS measurements improves the precision on the equation of state of dark energy by 15 per cent, to $w_0 = -1.010\pm-0.081$. 

\section*{Acknowledgements}

We are grateful to Florian Beutler, Antonio Cuesta, Daniel Eisenstein, Rita Tojeiro, and the anonymous referee for providing helpful comments that improved the quality of this paper.

AJR is thankful for support from University of Portsmouth Research Infrastructure Funding. LS is grateful to the European Research Council for funding. CH and AB are grateful for funding from the United Kingdom Science \& Technology Facilities Council (UK STFC). WJP acknowledges support from the UK STFC through the consolidated grant ST/K0090X/1, and from the European Research Council through grants MDEPUGS, and Darksurvey.

Mock catalog generation, correlation function and power spectrum calculations, and fitting made use of the facilities and staff of the UK Sciama High Performance Computing cluster supported by the ICG, SEPNet and the University of Portsmouth.

Funding for the creation and distribution of the SDSS Archive
has been provided by the Alfred P. Sloan Foundation, the Participating
Institutions, the National Aeronautics and Space Administration,
the National Science Foundation, the U.S. Department of
Energy, the Japanese Monbukagakusho, and the Max Planck Society.
The SDSS Web site is http://www.sdss.org/.

The SDSS I and II is managed by the Astrophysical Research Consortium
(ARC) for the Participating Institutions. The Participating
Institutions are the University of Chicago, Fermilab, the Institute
for Advanced Study, the Japan ParticipationGroup, Johns Hopkins
University, the Korean Scientist Group, Los Alamos National Laboratory,
the Max Planck Institute for Astronomy (MPIA), the Max
Planck Institute for Astrophysics (MPA), New Mexico State University,
the University of Pittsburgh, the University of Portsmouth,
Princeton University, the United States Naval Observatory, and the
University of Washington.

\label{lastpage}

\end{document}